\documentstyle[12pt,aasms4]{article}  
\def\msun{$M_{\odot}$}

\def\etal{{\it et~al.}~}
\def\hb{H${\beta}$~}
\def\ha{H${\alpha}$~}

\received{}
\lefthead{Stanghellini \etal}
\righthead{Magellanic Cloud Planetary Nebulae} 

\begin{document}

\title{{\it Hubble Space Telescope} Images of Magellanic Cloud 
Planetary Nebulae: Data and Correlations across Morphological Classes
\footnote{Based on observations made with the NASA/ESA Hubble Space Telescope,
obtained at the Space Telescope Science Institute, which is operated 
by the Association of universities for research in Astronomy, Inc., under
NASA contract NAS 5-26555} }

\author{L. Stanghellini\altaffilmark{2,3,4}}
\altaffiltext{2}{Space Telescope Science Institute, 3700 San Martin Drive,
Baltimore MD 21218, USA}
\altaffiltext{3}{Affiliated to the Astrophysics Division, Space Science
Department of ESA}
\altaffiltext{4}{{\it on leave,} Osservatorio Astronomico di Bologna}
\author{J. C. Blades\altaffilmark{2}}
\author{S. J. Osmer\altaffilmark{2,5}}
\altaffiltext{5}{Present address:
Department of Computer Sciences, University of Edinburgh,
King's Buildings, Mayfield Road, Edinburgh EH9 3JZ, Scotland}
\author{M. J. Barlow\altaffilmark{6}}
\altaffiltext{6}{Department of Physics and Astronomy, University College 
London, Gower Street, London WC1 6BT, UK}
\author{X.--W. Liu\altaffilmark{6}}

\newpage
\begin{abstract}

The morphology of planetary nebulae (PNe) provides an essential tool for
understanding their origin and evolution, as it reflects both the dynamics
of the gas ejected during the TP--AGB phase, and the central star
energetics. Here we study the morphology of 27 Magellanic Cloud planetary
nebulae (MCPNe) 
and present an analysis of their physical characteristics across
morphological classes. Similar studies have been successfully carried out
for galactic PNe, but were compromised by the uncertainty of individual PN
distances. 
We present our own HST/FOC images of 15 Magellanic Cloud PNe
(MCPNe) acquired through a narrow-band $\lambda$5007 [O~${\sc iii}$]
filter.  We use the Richardson--Lucy deconvolution technique on these
pre--COSTAR images to achieve post--COSTAR quality. Three PNe imaged
before and after COSTAR confirm the high reliability of our deconvolution
procedure.  We derive morphological classes, dimensions, and surface
photometry for all these PNe.  We have combined this sample with HST/PC1
images of 15 MCPNe, three of which are in common with the FOC set, 
acquired by Dopita \etal (1996), to obtain the
largest MCPN sample ever examined from the morphological viewpoint. By
using the whole database, supplemented with published data from the
literature, we have analyzed the properties of the MCPNe and compared them
to a typical, complete galactic sample. Morphology of the MCPNe is
then correlated with PN density, chemistry, and evolution.
\end{abstract} 


\section{Introduction} 

Planetary Nebulae (PNe) provide a fertile ground for
studying the evolution of low and intermediate mass (M$\le$8 \msun) stars.
The morphology of PNe, when combined with the physical properties of the
nebulae and the central stars (CS's), help to complete the picture of how
such stars evolve and how the evolution depends on mass, chemical content,
and PN environment. PNe are in fact stellar envelopes ejected in advanced
evolutionary stages, and carry a wealth of information on previous phases.

The morphology of PNe, as observed through narrow--band filters, traces
the structure of the ejected gas, and contains information on the
time interval between ejection and observation, in addition to the nature
of the ejection itself; the final ionized gas shape contains information
on inhomogeneities during ejection.  Morphological characteristics change
with both the nebular and stellar evolution, thus they carry a record of
the space and time history between the ejection and the observation. The
ejecta can be perturbed, for instance, by a fast CS wind, stellar
companions, planets in the central star's system, interstellar medium
condensations, magnetic fields, and by changes in the post--ejection stellar
evolutionary paths.

To date, several ground--based surveys of galactic PNe aimed at
delineating their morphological characteristics have been completed
(Schwarz, Corradi, \& Melnick 1992; Manchado \etal 1996; Chu, Jacoby, \&
Arendt 1987; Balick 1987). A space--based survey of galactic PNe has 
been performed
by Bond \etal (1995) with the WFPC2 camera on board the HST. Shapes of
planetary nebulae have been carefully classified and cross--correlated
with nebular and stellar properties, obtaining a series of interesting
results, ranging from the segregation of PNe hosting different types of
central stars based on their morphology (Calvet \& Peimbert 1983;
Stanghellini, Corradi, \& Schwarz 1993) to the indication that bipolar PNe
have more massive progenitors than elliptical PNe (Stanghellini \etal
1993; Corradi \& Schwarz 1994). However, such results need also to
be tested in a distance--bias
free environment.

Only a handful of galactic PNe have individually determined distances,
while the majority have distances derived with a statistical method, based
on physical assumptions, such as, for example, that all optically thin PNe
in the Galaxy share the same ionized mass (Cahn, Kaler, \& Stanghellini
1992; Kingsburgh \& Barlow 1992).  Not only are the absolute stellar and
nebular parameter determinations at risk when using poorly determined
distances, but even the morphologies themselves could be misclassified
when compared to one another and thought to be, for example, at the same
distance from us, since the detectability of morphological details
obviously decreases with distance. The proximity of the Magellanic Clouds
have made them perfect target galaxies to study PNe in a
distance--bias free environment. However, use of the cameras aboard HST is
required to resolve the shape of the Magellanic Cloud PNe (MCPNe). The
spatial resolution of HST/FOC allows observation of
MCPN morphology with similar definition to typical galactic PNe at a distance
of $\approx$
1.5~kpc observed from the ground with 1 arcsec seeing.

Only nineteen narrow--band HST frames of MCPNe have been published to
date, either imaged with the PC1 (Dopita \etal 1996, hereafter D96) or
the Faint Object Camera (Blades \etal 1992, hereafter B92). All published
data were acquired in the pre--COSTAR, pre--first refurbishing mission
epoch ($<$1994).  The B92 paper is an essay on what can be achieved with
HST/FOC when observing MCPNe through the \hb and [O~${\sc iii}$]
narrow--band filters.  B92 showed, for the first time, spatially resolved
images of four extragalactic PNe and their morphological details. The
target nebulae were chosen to be bright, so as to trace early post--AGB
evolution. 
Liu \etal (1995) used B92's images to construct detailed
photoionization models for two of these nebulae, SMC~N~2 and N~5, derived
nebular ionized masses and central star masses, and compared nebular
expansion ages with central star evolutionary track ages.  The main aim of the
GO observing program of D96 was to image, through the narrow--band
[O~${\sc iii}$] filter, a number of MCPNe covering a large domain in
nebular parameter space, such as [O~${\sc iii}$] luminosity, luminosity
class and optical thickness. Their published results included the
narrow--band images and the study of the expansion velocities and dynamical
ages, taking into account the nebular inclination on the plane of the sky
in the case of non--symmetric PNe. The resulting nebular evolution was then
coupled with a study of the evolutionary status of the central stars, by
means of the logT--logL plane location. Dynamical ages and evolutionary
times were found not to follow a simple correlation if the evolutionary
times were calculated on the basis of hydrogen-burning; rather, the
authors suggested in D96 that nebulae hosting helium--burning CS's 
outnumbered by
2:1 those hosting H-burning CS's.

The main purpose of the present paper is to present previously unpublished
narrow--band images of MCPNe, taken with the FOC as part of the
original FOC Investigation Definition Team science program. Most of the
images presented here were acquired before COSTAR was installed on the
HST. However, three MCPNe were re--observed following the first servicing
mission, using the FOC and COSTAR to check on the reliability of the
original images and the veracity of the deconvolution method used in their
analysis. We present the data as follows. In Section 2 we discuss the
observations obtained with the FOC, including target--selection criteria,
scheduling of the observations, data reduction and calibration, with
emphasis on the deconvolution method. Results from the FOC observations
are presented in Section 3 for the complete HST/FOC data set, i.e., the
newly observed PNe and the ones published by B92; we discuss [O~${\sc iii}$] 
images only. Additionally, we have broadened the
MCPN database by reclassifying the PC1 images from D96 with the same
morphological scheme used for the FOC data. So, in Section 3 we 
show (a) the final deconvolved
FOC frames, (b) the morphological classification, (c) the MCPN dimensions,
(d)  the photometry, and (e) the expansion velocity and the dynamical ages,
and discuss the results in 
Section 4. The conclusions, and a discussion of possible future developments,
are in Section 5.

\section{HST/FOC observations and reductions}

17 MCPNe have been observed since June 1991 over 4 HST observing cycles, using
the high resolution f/96 optical chain of the Faint Object Camera
(Macchetto \etal 1991).  The observations concentrated on using the
narrow-band F501N filter to record the [O~${\sc iii}$] lines, while a few
of the objects were also recorded in the narrow-band H$\beta$ filter
F486N.  The H$\beta$ recombination line traces emission by the dominant
element in the nebula, while the [O~${\sc iii}$] $\lambda$5007 transition
is a strong, collisionally excited, cooling line which traces emission by
the usually dominant O$^{++}$ ion of oxygen.  The systemic radial
velocities of the Small and Large Magellanic Clouds shift the 
[O~${\sc iii}$] lines by between +2 and +5~\AA\ relative to their rest
wavelengths, putting the $\lambda$5007 member of the [O~${\sc iii}$]
doublet at the peak of the filter's transmission. With a total bandpass of
74~\AA , the weaker member of the doublet at 4959~\AA\ will not be
transmitted through the F501N filter, according to the transmission curve
given in the current FOC Handbook (Nota \etal 1994).

This program was originally planned before the launch of HST, and the
targets were chosen to be easily detected through the selected filters
in reasonable exposure times and without any image saturation. The nebulae
were thus chosen to be reasonably bright in the $\lambda$5007 
[O~${\sc iii}$] line. In order to maximize the probability that the
chosen nebulae were optically thin, and thus that the ionized masses would
be equal to
the true nebular masses, two additional selection criteria were used: (a)
the nebulae should have a detectable $\lambda$4686 He~${\sc ii}$ line, a
standard criterion for selecting out low-- and medium--excitation PNe with
younger and less evolved central stars (Sanduleak, MacConnell, \& Philip
1978); and (b) the nebulae should have [O~${\sc ii}$] electron densities less
than about 5000 to 6000~cm$^{-3}$. The optical spectroscopic study of
Barlow (1987) had indicated this to be the dividing point between
optically thick and optically thin Magellanic Cloud PNe. The analysis by
Liu \etal (1995) subsequently confirmed that SMC N~2 with n$_e$(O~${\sc ii}$)
= 2850~cm$^{-3}$ and 3727/H$\beta$ = 0.29 is indeed optically thin but 
they found that 
SMC~N~5 with n$_e$(O~${\sc ii}$) = 3890~cm$^{-3}$ and  3727/H$\beta$ = 0.79
is still optically thick. 

Most of the observations were obtained in the observing cycles before the
1993 December servicing mission which repaired the imaging capability,
although three LMC objects, N~66, N~97, and N~192, have since been
re-observed.  As we shall explain, they provide an important check on the
analysis of the earlier data.  Because of its intriguing morphology, LMC
N~66 (SMP~83) was observed on two separate occasions before the servicing
mission as well as once afterwards. Our monitoring of this object turned
out to be fortuitous because of the recently announced variability in
brightness of the central star (Pena \etal 1994).  Although we use the
image of N~66 to check pre--COSTAR deconvolution performance, we do not
include this object in the discussion nor in Tables 2 and 3; Vassiliadis
(1996) has discussed and interpreted the ensemble of HST FOC and WFPC
images of N~66 obtained between 1991 July and 1994 February. 
%its peculiar shape and velocity structure deserve an individual study,
%which will be the object of a future paper. 
Table 1 gives a complete record of the FOC observations in the [O~${\sc iii}$] 
$\lambda$5007 light.  All observations
were taken in the 512$\times$512 pixel format with 25-$\mu$m square
pixels, corresponding on the sky to a plate scale of 0.0223 arcsec per
pixel before the servicing mission and 0.0144 arcsec per pixel with the
Corrective Optics Space Telescope Axial Replacement (COSTAR) in place
(Jedrzejewski \etal 1994).

The observations taken after the 1993 December servicing mission were obtained
very early in the observing cycle (January and February 1994) and before the
correct instrument sensitivities had been determined. Inappropriately,
pre--servicing values were in use at that time, resulting in an incorrect value
for the keyword PHOTLAM being attached to the data in the calibration pipeline.
 Unfortunately, all FOC data taken around that time have been archived with the
incorrect value.  (This problem was corrected in the FOC calibration pipeline
on 19 April 1994.)  We have rectified our Cycle 4 data with the proper values.
In order to guard against the risk of saturation to the post
servicing mission data we obtained a pair of F501N images for N~97 and N~192 by
taking a second image with a 1 magnitude neutral density filter, see Table 1. 

There are two calibrations that are applied to the raw FOC data in the
ground--system calibration pipeline, namely a geometric distortion correction
followed by a relative calibration or flat field correction. For the data taken
in the early cycles we reprocessed the observations using the IRAF/STDAS task
called CALFOC using new calibration files as they became available, and we have
reprocessed all the data originally presented in B92.  Finally, we removed by
interpolation the effects of the numerous reseaux marks, which are fiducial
reference marks engraved on the detector faceplate.

Our earlier analysis (B92) had shown that deconvolution techniques could be
employed to improve the qualitative appearance of these compact and high
contrast objects.  This result encouraged us to continue taking observations
throughout the early cycles rather than waiting for the 1993 servicing mission.
 Accordingly, we spent considerable effort in the reduction phase to try and
optimize the deconvolution and to investigate how different point spread
functions (PSFs) and the telescope focus were likely to be affecting the
accuracy of our data.  Subsequently, the three post--COSTAR images were of
great value in providing a direct comparison between the data sets and in
vindicating our approach. 

For deconvolution we used the non--linear restoration technique of Richardson
(1972) and Lucy (1974) which has been installed in the IRAF/STSDAS software
package, and we performed testing on the optimum number of iterations for our
images.  In qualitative terms, we found that after 50 iterations the images
showed considerable improvement, through reduction of the surrounding halo or
skirt of scattered light while still retaining the basic structure that could
be seen in the unprocessed images. After a larger number of iterations (100)
the shape and form of the objects began to break down and the images became
artificially lumpy and pixilated. We chose 50 iterations
for all the pre--COSTAR observations. The same number of iterations was found 
by Dopita and collaborators to be the best for deconvolving WF/PC images too
(Dopita 1998, private communication). 

The images obtained in 1994 with COSTAR of LMC N~97 and N~192 were valuable in
establishing the veracity of the deconvolution work.  In Figure 1 we show contour
plots of the deconvolved pre--COSTAR images of these two PNe with the more
recent images obtained with FOC and the COSTAR. There is
excellent agreement between the pre--COSTAR deconvolved images and the
post--COSTAR direct images. The agreement covers the overall size and shapes
of the nebulae and extends to the smallest structures that can be discerned at
a scale of 0.07 to 0.1 arcsec. The consistency can also be seen in the 
images that are presented later (see Figure 5, panels 1 and j, and m and n), 
as well as in the good photometric agreement (see $\S$ 3).
The direct comparison provides confidence that the deconvolution
technique has improved the qualitative appearance of these objects without
introducing artifacts. 

Routine monitoring of the image quality of HST has been carried out since
launch in order to monitor and maintain good telescope focus as well as to
characterize the features of the optical performance (Hasan \& Burrows 1994). 
During this time, the Optical Telescope Assembly (OTA) continued a steady
contraction as gas desorbed out of the telescope structures, thus requiring
frequent re--alignments to retain the focus within 10 microns of the nominal
value, and this was not always achieved.  In addition, short time period
fluctuations of the OTA PSF were discovered (Hasan \& Bely 1993) which are
attributed to expansion and contraction of the secondary mirror support system
causing small ($\approx$5 microns) motions of the secondary mirror (breathing).
 Also, the internal focus of the FOC was optimized on occasions. It is
impossible to unravel the effects of these optical changes from our
observations because simultaneous PSF observations were not obtained. (Indeed,
it would have been time consuming to have attempted to calibrate these
defocusing effects, especially the breathing which can alter the PSF by small
amounts over short time periods of about 30 minutes.) 

We were concerned that these image degradation problems could affect the
resolution of our data and we ran tests to see how sensitive our deconvolution
results were to different input PSFs.  We experimented with a variety of
Richardson--Lucy deconvolutions. Observed PSFs were obtained from FOC
calibration observations of BPM 16274 through the F501N filter and these
observations provided two PSFs, one based on observations from early in Cycle 1
and one from Cycle 3.  A search of the FOC archive yielded a third PSF star
from April 1992 observations of the SN~1987A field.  Finally, we constructed a
theoretical PSF using the optical modeling work of Krist (1993).  An obvious
advantage of the theoretical PSF compared with any of the observed PSFs is the
infinite signal--to--noise ratio.  Figure 2 shows the four PSFs described above
and Figure 3 shows the results from deconvolving the pre--COSTAR image of N~192
with each of these PSFs. 

As inspection of Figure 3 confirms that, at the level to which we are working,
the Richardson--Lucy deconvolution is not very sensitive to the input
PSF.  Both the overall shape and most of the small scale structures remain
the same over all four images.  On closer examination there are subtle
changes from one deconvolution to another, and these changes provide an
empirical assessment of the overall accuracy of the deconvolution work. 
In general, high signal--to--noise ratio PSFs gave the best results
(higher counts in the peak of the PN image).  This seemed a more important
parameter than the closeness in time of the PSF to the actual observation. 
In other words, for our data, a PSF produced from an observation in 1991
worked at least as well as a PSF produced in a 1993.  Probably the
breathing phenomenon is destroying any advantage the contemporary PSF may
otherwise have had.

The theoretical PSF can be adjusted to fit precisely any observation by
matching the Airy ring pattern using a stellar image in the field of the PN.
Potentially, this could allow correction for the breathing phenomena.
Unfortunately, among our images only that of N~66 has a star (in fact the
central star) suitable for such matching.  Figure 4 shows the results of
matching and not matching the theoretical PSF to the central star.
Qualitatively it is difficult to discern the difference between the two;
however, the peak counts in the image deconvolved with the matched Airy rings
are 20 percent higher, indicating that an improvement is achieved with this
method.  On the other hand, we found that use of the theoretical PSF tends to
yield a rather uneven and blotchy appearance to the image which the observed
PSFs do not.  In any case, the lack of available stars in the other PN images
prevents us from using this approach for cases other than N~66.  After
considerable testing we selected the 1991 observed PSF to use for all our
pre--COSTAR denconvolution work. 

A particular concern that we had was whether faint extended halos that
might exist around the main nebular structures would be recovered by the 
deconvolution process, since such halos could potentially contain
a significant fraction of the total nebular mass. For example, 
a faint halo three times larger than that of the
main inner structure, and with a mean surface brightness of only 1.5\%
that at the central peak, would still contain as much mass as in the inner
bright nebula. We therefore experimented by artificially adding smooth
halos of varying surface brightnesses and diameters to the deconvolved
FOC [O~${\sc iii}$] nebular images of SMC~N2, SMC~N5 
and LMC~N192. These composite images were then convolved with
an observed PSF and the resulting convolved images were deconvolved
with the Richardson-Lucy algorithm in the standard way. It was found
that the artificial faint halos were recovered in the deconvolved images
in all cases, down to halo surface brightness levels of 1\% of the
peak inner emission. We are therefore confident that any halo emission
around the brighter PNe in the sample
must be below this level, although for the fainter PNe in the
sample the upper limits to any surrounding halos would be 
correspondingly larger. Our re-observations
of four of the nebulae, after the installation of the COSTAR corrective
optics, revealed no extended low surface brightness emission around
them, confirming this conclusion in their cases.

\section{Image analysis} 

\subsection{The FOC and PC1 data sets} 

Our discussion of PN morphologies, dynamical expansion times, and
luminosities is based on two data sets. The first data set is composed of
the FOC images presented here for the first time and the images
illustrated by B92 (FOC set); the other set consists of the PC1 images
described by D96 (PC1 set). In all we have 27 MCPNe, excluding N~66
(see $\S$ 2).
The list of
observed PNe can be found, respectively, in Table 1 of this paper for the
FOC set, and in D96's Table 1 for the PC1 set. 
The two sets have three objects in
common, useful to check the criteria for morphological classification, and
image quality. Each set has been selected with a defined criterion: the FOC set
contains PNe with high [O~{\sc iii}] fluxes and generally low optical
thickness, while PC1 set presents a variety of [O~{\sc iii}] fluxes and
Lyman continuum optical depths. As a consequence, the two sets are not
homogeneous, and the final composite sample is not, by any means, a
complete or unbiased statistical sample of MCPNe. 
Nonetheless, there is purpose to analysing the
composite group of PNe in a qualitative way, in order to 
establish morphological trends.

\subsection{Morphology and diameters}

The $\lambda$5007 [O~${\sc iii}$] narrow--band FOC images are presented in
Figure 5.  The pre--COSTAR images are deconvolved, as discussed in Section
2. In Figure 5 we also include three nebulae already published by B92,
and we show both the pre--COSTAR (deconvolved) and post--COSTAR images 
for N~97 and N~192. 
The following discussion, on the morphology, the dimensions, and the
photometry of the PNe observed with the FOC, is based on the 
[O~${\sc iii}$] images. To classify the morphology we follow the most recent 
and widely used scheme by Schwarz, Corradi, and Stanghellini (1992) in its
updated version (Manchado \etal 1996). Originally, this classification
scheme was conceived for \ha (or \hb) images, expecting this emission line
to track the bulk of ionized gas in most PNe. In the case of the FOC set, whose
PNe have high excitation, the morphological differences between high--and
medium--excitation plasma tracers are not expected to be significant (for
galactic equivalents, browse through high--excitation PN images in the
catalog by Manchado \etal 1996). In the case of the PC1 set, however,
the lower excitation might not be fully delineated
by the [O~${\sc iii}$] line.
Another source of inhomogeneity among the two data sets is the different
angular resolution of the two cameras used in the observations. The PC1
frames published by D96 clearly show their lower angular resolution with
respect to the FOC images published here for the first time, or by B92. 

The classification scheme sorts PNe into five main groups, as defined by
the outer envelope of the PNe: Round PNe (R), Elliptical PNe (E), Bipolar
PNe(B), Quadrupolar PNe (Q), and Pointsymmetric PNe (P). Bipolar PNe are
nebulae with one axis of symmetry, and a detectable {\it waist}.
Quadrupolar PNe consist of two pairs of bipolar lobes, joined at a common
waist. Pointsymmetric PNe show structures that are symmetric with respect
to a central point (in 2D). The scheme by Manchado \etal (1996) does not include
``irregular PNe'', although it includes the possibility that a PN could
not be classified within the above scheme, and in this case they are
denoted as NC. The main classes have subclasses, denoted by suffixes
attached to the morphological main symbols. The subclasses describe inner
structures (s), multiple shells or haloes (m), ansae attached to the main
structures (a), and rings at the waist of some bipolar nebulae (r).

We can apply the morphological classification scheme to Magellanic Cloud
PNe, although we should keep in mind that the [O~${\sc iii}$] images of
these nebulae will track the bright cores rather than outer features, such
as multiple shells and bipolar/quadrupolar lobes. In this sense, real
bipolar structures may not be observed in their complete display of lobes,
rather only the inner ring may be visible. As discussed in Section~2, our
analysis of the nebulae confirmed the absence of lobes.
We find instead a considerable subgroup of objects whose outer shape is
elliptical, and whose inner shape is ``bipolar'', showing two
concentrations of surface brightness. Such structure is reminiscent of a
projected inner ring, that, in turn, is typical of bipolar outflows. From
the asymmetry of the ring like structures we can be quite confident that
their true morphology is bipolar rather than elliptical. We classify these
PNe as {\it bipolars} (B), subclass {\it bipolar core} (bc).

Table 2 gives the morphological classification of PNe. In columns (1) and
(2)  we give the discovery name and SMP (Sanduleak \etal 1978) catalog
number.  Column (3) gives the HST camera used for the observation, where 
{\it FOC} indicates the PN whose
images are published in this paper for the first time, or by B92, and 
{\it PC1}
indicates the D96 images. Column (4) defines the morphological class, and the
detailed classification in parenthesis. In two particular cases, the shape
is incomplete, and we define that particular shape as the suffix `inc'. In
column (5) and (6) of Table 2 we give the angular and physical diameters
of the PNe, respectively in arcsec and parsecs.

Our diameter measurements are a result of a detailed photometric 
routine, as we describe following.
First, we define a geometrical center of the
PN on the image, by hand. Second, we chose for each object a set of (circular) apertures that segment the nebula into anuli. The outermost of
these apertures is set at a large distance from
the apparent nebular limb. Then we measure the flux in each
aperture with the IRAF/PHOT routine,
obtaining the sky--subtracted total flux within each aperture. 
Going outward
from the center, we find the maximum total
nebular flux. We then plot the relative encircled flux (flux within 
each aperture divided by the total flux) versus aperture for each nebula,
and we read out the aperture encircling 85 \% of the total
flux. We define the latter aperture to be the physical nebular radius.
This procedure has been repeated for pre-- and post--COSTAR images for N~97
and N~192, obtaining satisfactory agreement.

Among the 15 diameters measured with the photometric method by us,
six objects show the presence of a ring in the relative encircled flux
profile: N~4 (Bbc), L~305 (Es), L~536 (E), L~343 (Bbc), N~18 (Bbc), and 
N~67 (Bbc). This photometric check is a good method to determine which PNe actually show ring--like features.
By using the above described method,  
we had difficulties in finding the outer contour of the
nebula N~24, or the size at which its encircled flux become constant.
This planetary has a halo/core structure, with FWHB of the bright core 
measuring 0.23 arcsec. We thus do not give is outer dimensions in Table 2,
and we eliminate this object
in the discussion of the results and in Figures 6 through 10.
Future observations of this particular object are in order.

Our definition of physical radius is, in principle, the same as D96's,
and has been chosen this way for uniformity among the
two data sets. Nonetheless, of the three objects in common among the two sets,
only for one (WS~12) do the two measurements agree. In the cases of N~4 and
L~536, both with ring--like profiles, the measurements are different.
The difference can be ascribed to the power--law skirts produced by
incomplete deconvolution of the PC1 set, and adds an extra uncertainty to our discussion.
For the PC1 set we derived the angular sizes from the published physical
sizes (see Table 3 in D96) and the distances to the Clouds quoted therein
(d$_{\rm LMC}$=50.60 and d$_{\rm SMC}$=58.29 kpc). Obviously the same distances
to the Clouds have been used to derive physical sizes for the FOC set.

Below, we explore in some detail the individual morphologies of those
nebulae whose structures are not spherically symmetric. For PNe showing an
asymmetric structure, we estimate the projection angle on the plane of the
sky, measured from the ratio of the semiminor to the semimajor axes of the
ring--like structures, and we have included this axial ratios, q, in
Table 3.  

\noindent{\bf N~2} is a regular ellipse with an inner hole. 

\noindent{\bf N~4} has an elliptical, {\it boxy} shape with evidence of an
inner, projected ring.
D96 classified it as {\it BR} (bipolar/ring), which,
apart from the different terminology, corresponds to our definition. 

\noindent{\bf N~5} is almost round, with an inner hole. 

\noindent{\bf N~18} has a fairly round outer shell, and an inner, edge--on ring.

\noindent{\bf L~305} is elliptical; contour levels show a marked asymmetric
structure in the inner parts, as if the maximum brightness was off--center. 

\noindent{\bf N~67} resembles a ring feature of a galactic bipolar PN (e.~g.
NGC~650, Sh~1--89, Manchado \etal 1996); the measured inclination of the
ring is approximately 45 degrees on the plane of the sky. 

\noindent{\bf L~343} is elliptical with a ring--like core. 

\noindent{\bf L~536} is elliptical, with low ellipticity. D96 defined 
it as {\it s} (spherical), while we can actually measure an axial ratio of
about 0.8.

\noindent{\bf LM2--5} is elliptical, with an asymmetric ring--like core. 

\noindent{\bf N~97} shows four density enhancements, and can be classified
as quadrupolar. 

\noindent{\bf N~24} has a very slightly elliptical outer shell, and a regular
round inner shell, and can be classified as R. 

\noindent{\bf N~192} has a slightly elliptical outer shell, an irregular inner
structure, and presents an inner hole. It is classified as R.

\noindent{\bf WS~12} has elliptical contours, with an incomplete
crescent--shape structure. Although D96 classified it as {\it BR}, we
could not definitely see the complete ring with our FOC image brightness
analysis. 

\noindent{\bf WS~16} is a ring--like structure, of elliptical contour; we
classify it as Es. 

\noindent{\bf LM1--27} has an irregular inner structure, reminiscent of an
incomplete ring. 
 
\noindent{\bf N~122}: although D96 found a bipolarity on the deconvolved image,
we could not find it on their raw image, which, on the contrary, shows a
genuine elliptical PN, with very high ellipticity.

\noindent{\bf N~52} is elliptical with an inner hole. 

\noindent{\bf LMC SMP~72} is very hard to classify. At a first glance it
could resemble a quadrupolar, but a careful analysis shows no evidence for
the second pair of rings. It can be a bipolar with an enhanced, large
ring.

\noindent{\bf N~60}: our morphological classification confirms that of D96,
of shperical/round shape.

\noindent{\bf N~215} is elliptical with a bipolar core. 

\noindent{\bf LMC SMP~96} is elliptical with a bipolar core. 

\noindent{\bf LM1--61} is round, with irregular inner brightness. 

\subsection{Aperture photometry of FOC images} 

Aperture photometry has been performed for the FOC images. Calibrated, but
non--deconvolved, images were used to this end.
Even if most PNe are easily contained
in a 2$\times$2 arcsec$^2$ aperture, we chose an aperture of 6 arcsec$^2$,
since pre--COSTAR images may contain considerable energy output out to 4
arcsec from the target centers. We measured the total counts per second
within the aperture, after sky subtraction (Tab.~1, col.~[8]), the peak
counts 
per second (Tab.~1, col.~[9]) and the
calibrated physical fluxes (Tab.~1, col.~[10], in erg cm$^{-2}$ s $^{-1}$).
The derived FOC [O~${\sc iii}$] $\lambda$5007 line fluxes show 
excellent agreement with the ground-based values measured
by Jacoby, Walker, \& Ciardullo (1990). For 21 FOC
measurements, the mean flux difference is found to be just 
0.00$\pm$0.02 dex.

\subsection{Expansion velocities and dynamical expansion ages} 

In order to evaluate the dynamical expansion ages of our PNe we
need their expansion velocities. 
%For six of the PNe that we analyzed we use data acquired in 1994 with the 
%Manchester Echelle Spectrograph mounted on the 3.9m telescope 
%at the AAT Observatory. The echelle velocities used here have never been
%published before, and a more extensive analysis of the echelle data will 
%be the subject for a future paper. The spectra were analyzed via 
%multi--gaussian analysis, to evaluate the (possible) multiple velocity 
%components. For the remaining PNe, 
We have used velocities based on measurements published by Dopita
\etal (1985 for SMC, 1988 for LMC PNe). 
%Both Dopita \etal's papers are based on data collected with the 2.3m 
%telescope of Siding Springs Observatory, with lower signal--to--noise 
%ratio than the AAT. When using the two sets of velocities (AAT and MSSSO) 
We should beware that Dopita
\etal define the expansion velocity as 0.911 times the FWHM of the
$\lambda$5007 [O~${\sc iii}$] line, corrected for instrumental and thermal
broadening, whereas in general (e.g. in the galactic PN expansion
velocity catalogs of Sabbadin 1984 and Weinberger 1989) the value v$_{\rm
exp}$=0.50 FWHM is used for unresolved nebulae\footnote{Robinson, Reay \&
Atherton (1982) have shown theoretically that the FWHM linewidth of a
nebula
completely enclosed by an observing aperture is equal to the line
splitting that would be observed at the nebular center in a spatially resolved
observation. Munch, Hipplelein, \& Pitz (1984) have confirmed this result
observationally}.  We use this second choice for the expansion velocity,
and corrected the velocities of Dopita \etal as if
they have been measured in this way, thus dividing them by 1.82.
The
resulting nebular expansion velocities are given in Table 3, column (3)
\footnote{
The FWHM of a Gaussian line profile contains 76\% of its total
flux.  Dopita \etal (1985) defined the expansion velocity as the
half-width at one tenth maximum line intensity. For a Gaussian, the full
width corresponding to this definition contains 97\% of the total line
flux. For comparison, the nebular diameter definition adopted by D96
and by ourslves is the diameter encircling 85\% of the total
nebular flux. For a Gaussian line profile, 85\% of the total line flux is
contained with 0.4 maximum line intensity. We prefer to adopt here the
usual definition of v$_{\rm exp}$ = 0.5~FWHM, but if expansion
velocities coresponding to the half width at 0.4 maximum line intensity
are preferred, then the derived expansion ages in Table~3 should be
decreased by a factor of 1.15.}.

Column 4 of Table 3 lists $\tau_{\rm dyn}$ = R$_{\rm neb}$/v$_{\rm exp}$, 
the dynamical expansion ages derived from the PN radii and expansion
velocities (where R$_{\rm neb}$ is half the diameter D listed in col. 6 
of Table 2). Table 3 (col. 2) also lists q, the measured ratio of the nebular 
semi-major to semi-minor axes. D96 made use of this parameter to 
correct dynamical expansion times for nebular inclination effects, 
assuming that ring-shaped nebulae are circles viewed at an
inclination angle $\theta$ = cos$^{-1}$q with respect to the
plane of the sky, so that measured expansion velocities should
be corrected for inclination effects by dividing them by sin $\theta$.
However, since this would yield infinite expansion velocities and
zero expansion ages for q = 1, no correction was made for circular
nebulae. We found that the use of this scheme led to large
decreases in the derived expansion ages for nearly-circular nebulae
(e.g. a factor of 2.5 for SMC N2, with q = 0.92), versus no correction
at all for perfectly circular nebulae (e.g. SMC N5, q = 1.0) and so decided not
to make such a correction. We note that for non-circular nebulae the 
nebular radius defined by the 85\% encircled energy definition is in any
case a mean of the semi-major and semi-minor axis dimensions, so that  
its use yields dynamical ages that are smaller than those that would be
obtained just from the semi-major axis dimensions. We note that barrel-shaped
nebulae can yield apparent circular shapes when viewed pole-on, and
elliptical shapes when viewed equator-on. Figs. 4 and 5 of Frank \&
Mellema (1994) show that for such nebulae viewed pole-on the measured expansion
velocity corresponds to material along the line of sight that is expanding
in the polar (longer axis) direction, with a velocity higher than that
in the equatorial direction. Thus dynamical ages for apparently near-circular
nebulae of this type may therefore be underestimated, since they could
be using too high an expansion velocity.

\section{Analysis of the results} 

The morphologies of the 27 MCPNe in the [O~${\sc iii}$]
narrow band images are similar
to those of galactic PNe, if we consider the bright parts of the latter
ones. We did not find multiple shell PNe or faint extended lobes of
bipolar and quadrupolar PNe. Similarly to galactic PNe, we encounter
round, elliptical, bipolar (ring), and quadrupolar shapes. We did not
expect that the statistical distribution among our group of MCPNe would be
the same as for galactic PNe, since we have overall selected against faint
and low excitation PNe, thus against symmetric shapes (Stanghellini \etal
1993). We found that 36\% of the studied MCPNe are round, 32\%
are elliptical, and 32\% have bipolar or quadrupolar shapes. The
northern galactic sample (Manchado \etal 1996) has 24\% round, 56\%
elliptical, 17\% bipolar and quadrupolar, and 3\% pointsymmetric PNe.
We thus confirm the existence of three main morphological classes, round,
elliptical, and bipolar/quadrupolar PNe. We did not find pointsymmetric
PNe, nor did we expect them, given the low percentage of occurrence of
this particular morphology among galactic PNe. We confirm that more
bipolar PNe can be found among high excitation objects, as was already
inferred from Zanstra analysis by Stanghellini \etal (1993). Our
statistical analysis cannot proceed any further, given that we do not have
a statistically significant sample.

The main advantage of studying MCPNe with respect to their galactic
counterparts resides in knowing their distances. Distance--dependent
physical properties, such as physical dimensions, dynamical times, and
luminosities, are readily determined for MCPNe. When we discuss dynamical
times derived from physical dimensions, we should not overlook the fact
that some nebulae are optically thick to the ionizing radiation from the
central stars. If a PN should remain optically thick for most of its
evolution, its measured diameter would not trace the dynamical evolution,
but rather the evolution of the ionization front. We have sorted our PNe
according to their optical thickness, as derived from the line ratio
$\lambda$3727 [O~${\sc ii}$] / \hb. As Kaler \& Jacoby (1990) pointed out, 
this ratio should be higher than 0.8 and 0.35 for, respectively, LMC and 
SMC PNe to be optically thick. We derive the diagnostic ratio from 
spectral line intensities available in
the literature (Meatheringham \& Dopita 1991ab; Vassiliadis \etal 1992), 
and report the optical thickness in Table 2, Column (7).
This measure of thickness is rather crude, in that it does not take into account
variations of the diagnostic ratio with density, thus a small fraction of
PNe labeled 
as {\it thin} in Table~2 might be in fact thick. The results of Table~2
agree for the most part with Dopita and Meatheringham's
(1991ab) photoionization models optical thickness, which we do not use therein 
to avoid model dependence. 

Among those PNe whose diagnostic spectral lines are available in the 
literature,
we find that (a) about half the 
elliptical PNe are optically thin, (b)
most round PNe are optically thin according to the above
criterion, and (c) only one asymmetric (bipolar) PN
is optically thin. Obviously the fact that the majority of PNe in the FOC set
are optically thin to the ionizing radiation strongly depends on the target
selection of those planetaries, but the thickness/thinness of each 
morphological class
was not selected a priori. Since most bipolar/quadrupolar PNe are thick to ionizing radiation, their measured physical size can be
an underestimate of the real size, and the dynamical time could be
actually larger than calculated.

In Figure 6 we plot the histogram distributions of three main nebular
properties: physical dimensions (top), expansion velocities (middle), and
dynamical expansion ages (bottom). Each morphological class is represented
in a different way (see caption).  we infer the following properties: (a)
bipolar PNe have dimensions larger than 0.2 pc, this result, although
based on very few objects, is an important confirmation of a similar
situation existing for galactic PNe (Stanghellini 1995); (b) bipolar PNe
in our sample have physical dimensions within a narrower range than
elliptical and round PNe.  

In Figure 7 we examine the time evolution of the PN sizes for three major
morphological classes: round, elliptical, and asymmetric (bipolar and
quadrupolar) PNe. We did not include those PNe whose angular size is a
measured upper limit (see D96). 
The physical dimensions correlate linearly with the
dynamical ages, as expected, with scatter due to the velocity
distribution. In particular, elliptical and bipolar/quadrupolar PNe 
define a very tight correlation, with coefficient R$_{\rm xy}$=0.92. 

We can use the physical dimensions and dynamical age as independent
variables to reveal correlations with other physical parameters across
morphological classes. Due to the limited size of our sample and the
selection criteria of the targets, the range of physical diameters is
rather restricted, and
each morphological class is not statistically represented. When using the
dynamical age as an indication of the evolutionary timescale, we should not overlook the fact that it merely indicates the time lapse between the
envelope ejection at the TP--AGB phase and the observing time, and assumes
a constant expansion velocity without acceleration (or deceleration) or
the shell. $\tau_{\rm dyn}$ is a very useful variable for order of
magnitude correlations, but it does not indicate the exact lifetime of a
PN. Furthermore, since zero age post--AGB tracks generally correspond to a
defined central star temperature, one can really never finely tune these
tracks to the observed dynamical times, and a direct comparison among the
two sets of parameters, the theoretical ones and the empirical, should not
be used without precautions (K\"aufl, Renzini, \& Stanghellini 1993). 
On the other hand, dynamical
ages measured for MCPNe are generally more homogeneous than those measured
for galactic PNe since their distances are better known and the dimensions
and the velocities of the MCPNe both correspond to the high excitation
body of the PN.

Electron densities, measured from forbidden line ratios, have been plotted
in Figure 8
against the physical dimensions of the MCPNe.
The general trend shows a decreasing electron density with increasing 
physical size, with the exception of L~305 and N~67, whose loci are
in the upper--right part of the diagram.

In order to study the fading of PNe  with evolution, accordingly
to their shapes, we have
analyzed the [O~${\sc iii}$] surface brightness. 
The [O~${\sc iii}$] luminosities from which we derive the surface brightness
have been
calculated from the total fluxes observed form the ground
(Jacoby \etal 1990), the Cloud distances, and the extinction
constant (Boffi \& Stanghellini 1994, and references therein). 
The correction for extinction has been performed using the galactic extinction
curve (Osterbrock 1989), which around 5007~\AA~ has a similar shape to 
the curve derived for the
Magellanic Clouds (Hoyle \& Wickramasinghe 1991). Figure 9 aims at
disclosing possible evolutionary effects on the
surface brightness
for PNe of different shapes. The [O~${\sc iii}$] luminosity depends on the
stellar energetics, and secondarily on the oxygen content and on the
effects of nebular evolution (Richer 1993). It is thus a good guide for tracing the intrinsic stellar luminosity.

What we see in plot 9 is that the round PNe (symbols are as in the other Figures) are not to be found at low surface brightness, as opposed 
to elliptical/bipolar/quadrupolar PNe. 
One reason for the split in fading behaviors could certainly be a difference
in the ionized masses, which, in turn, could be an indication for a difference 
in the mass of the progenitors (see Fig.~8 in  Boffi \& Stanghellini, 1994). 
On the
other hand, a systematic difference of velocity fields among the round PNe
and all other shapes can also produce such a separation as in Figure 9.
Indeed, the relation between the axial ratio q and v$_{\rm exp}$ shows 
that extreme
asymmetric PNe do evolve faster. 
But the velocity difference alone does not explain the discrimination among 
morphological types of Figure 9. Unfortunately, given the small size of 
the sample, and that most of the PNe in the Figure are optically thick
 to ionizing radiation, we can not conclude that
we are
observing two groups of PNe with different progenitor masses. 
What we are probably seeing here is that more massive stars evolve faster
through the high luminosity post--AGB phase, and they are fading at the
time of observation. On the other hand, stars with low--mass progenitors
evolve slowly, thus retaining their high luminosity for a longer time.
Should this interpretation be right, we can conclude that most
bipolar/quadrupolar and elliptical PNe in our sample have high mass 
progenitors, while
most round PNe have low--mass progenitors, in addition to lower velocities.
Only with a much larger and homogeneous sample of MCPNe we could investigate
this important aspect of PN evolution to its fullness.
Accurate modeling of the surface brightness evolution of
expanding shell and ring PNe are also required to provide the necessary 
background to complete the picture.

In order to confirm the nature of the asymmetric PNe in our sample, and
to test the correlation between morphology and chemical enrichment, as
found by Peimbert and collaborators (e.g., Calvet \& Peimbert 1983, Peimbert 
\& Torres-Peimbert 1983),
in Figure 10 we plot the N/O abundance ratio against
dynamical expansion time for all PNe for which the N/O ratio is available
(Richer 1993). 
Symbols are for the different nebular shapes, as in the other Figures.
Since the (revised) N/O
abundance ratio constraints for Type I PNe are different for SMC, LMC and
Galactic PNe (Kingsburgh \& Barlow 1994, Peimbert 1997), we have
artificially decreased the N/O abundance ratio of SMC PNe by an
appropriate factor, so that they can be directly compared ($\Delta$ log
N/O=0.24). The horizontal line in Fig.~10 is at the appropriate level so
that PNe above the line are of Type I (see note {\it a} on Table~2 for 
PNe identification). We find that all round PNe in our
sample are non--Type I, all but one
bipolars are Type I, and elliptical PNe are
equally divided among the two Peimbert Types. We do not see any evolution
in the N/O abundance. The number
of objects is so low to leave the possible consequences unexplored for
now. What we can infer from the last two Figures is that round and
bipolar/quadrupolar PNe form two distinct ``enrichment'' groups, while more
investigation is necessary to determine whether the ellipticals are an
intermediate sequence or a different evolutionary stage of either
round or bipolar nebulae. 

\section{Conclusions} 

We have presented a set of narrow--band images of 15 Magellanic
Cloud Planetary Nebulae acquired with the FOC, aboard the Hubble Space
Telescope.  Deconvolution techniques, and comparison to post--COSTAR FOC
images of three of the PNe, show that excellent image quality can be
achieved from pre--COSTAR images.  
We have measured the nebular angular diameters, allowing the calculation
of dynamical expansion ages, when combined with the known distances to the
Magellanic Clouds and previously measured expansion velocities. We have
used the published PC1 MCPN images from D96, and other relevant
physical parameters from the literature, to obtain a total group of {\it
27 extragalactic planetary nebulae with known distance, morphology, and
dynamical age}, by classifying all the PNe with the same morphological
scheme. 
The main scientific content of this paper is the presentation,
discussion, and
analysis of the new MCPN data acquired with HST/FOC.
We also attempted a limited analysis of nebular properties across
morphological classes. The results suffer from low statistics, especially
within each class. Nonetheless, we find some trends that would confirm previous
studies on galactic PNe, mainly, that symmetric and asymmetric PNe seem to
belong to different brightness group (in the $\lambda$5007 [O~${\sc iii}$] line), possibly
indicating that they belong to different mass groups. In order to have
greater consistency in these results, we would need to discuss at least 20
objects for each morphological class, that is, a sample of a
hundred MCPNe observed with the HST cameras. 
Moreover, although the morphological classification is feasible using pre--COSTAR images, the photometric measurements can suffer 
considerable errors; it is thus necessary to repeat and extend the analysis
to post--COSTAR images of MCPNe. 
More insight into the
evolutionary paths of different nebular shape classes could be achieved by
investigating the morphological properties of the MCPNe together with
their central stars. The analysis of the sample of MCPNe presented in this
paper, together with their central stars, is in progress and will be
published in the future.

\acknowledgements 

Thanks to D.~Shaw for discussions on the correlation between nebular
morphology and stellar evolution in the Clouds, and
to M.~Dickinson for helping with the IRAF routines.
The referee of this paper, M.~Dopita, is thanked for his insight and 
very useful comments.
L.S. gratefully acknowledge the hospitality at the Space Telescope 
Science Institute, where this work was completed. J. C. B. acknowledge support 
from NASA through the contract NAG5-1733.

\clearpage

\centerline{\bf Figure Captions}

\noindent {\bf Figure 1}. Contour
plots of the LMC N~192 (a, b) and N~97 (c, d) pre-- and post--COSTAR images.
 
\noindent {\bf Figure 2}. 
Various point spread functions (PSFs).

\noindent {\bf Figure 3}.
Deconvolution of the pre-COSTAR image of N~192 by using
each of the PSFs in Fig.~2. 
 
\noindent {\bf Figure 4}. 
Results of matching and not matching the theoretical PSF to the central
star, in N~66.

\noindent {\bf Figure 5}. 
Narrow--band [O~${\sc iii}$] FOC images of Magellanic Cloud
planetary nebulae.

\noindent {\bf Figure 6}.
Parameter distribution of nebular size in pc. (top),
nebular expansion velocity in km s$^{-1}$ (middle),
and dynamical expansion time in yr (bottom panel) for round (solid),
elliptical (dashed) and bipolar/quadrupolar (shaded histogram) MCPNe.

\noindent {\bf Figure 7}. Maximum nebular dimension versus dynamical 
expansion age for MCPNe which are round 
(open circles), elliptical (solid circles), and bipolar/quadrupolar
(squares).

\noindent {\bf Figure 8}. Nebular electron density versus physical
dimension for PNe. Symbols are as in Figure 7.
Diameters of optically thick PNe, and of PNe with unknown thickness,
can actually be larger than observed, and indicated with arrows.

\noindent {\bf Figure 9}. [O~${\sc iii}$] surface brightness
versus dynamical
expansion time. Symbols are as in Figure 7. 

 \noindent {\bf Figure 10}. Nebular N/O abundance ratio versus dynamical
expansion time.
Symbols are as in Figure 7. The solid line represents the
dividing line between Type I (top quadrant) and non--Type I PNe (bottom
quadrant), as described in the text.

\clearpage

\begin{deluxetable}{l c c c c c r c c}
\scriptsize

\tablecaption{
FOC F/96 Observations of MCPNe in [O~${\sc iii}$]}

\tablehead{
\colhead{Name}&    \colhead{RA\tablenotemark{a}}   &     
\colhead{Dec\tablenotemark{a}}    &    
   \colhead{Filter}  & \colhead{Date}  & \colhead{t$_{\rm exp}$} &                     \colhead{counts}   &   
\colhead{peak counts} & \colhead{log F} \nl
& (J2000)  &(J2000)&   &     
 (UT)  & [s]    &    [s$^{-1}$] & 
[s$^{-1}$]& [erg cm$^{-2} s^{-1}$] \nl}

\startdata

\cutinhead{SMC}
N~2    &    00:32:38.8&    --71:41:59&     F501N & 1991
Jul 09.62   &   995.9   &    407.27  & 0.442 &  --11.70\nl  
N~4   &  00:34:22.0   &  --73:13:21   &   F501N   &  1993 Apr 28.92   &496.8   &     137.05& 0.156& --12.19 \nl  
%
%N~5   &  00:41:21.8   &  --72:45:19   &     F486N    &   
%1991 Jul 9.82   & 921.4    &         56.30   &   --12.810    &   0.056   
% &   28.13\nl
%    
N~5&  00:41:21.8  &  --72:45:19&    F501N   
 &   1991 Jul 09.76   &   995.9   &  334.65 &   0.413  &  --11.79\nl  
N~18   &  00:46:59.6   &  --72:49:39   & 
 F501N   &  1992 Nov 25.31   & 996.9   &      154.07  &  0.168  & --12.14\nl   
L~305   &   00:56:30.9   &   --72:27:01  
&    F501N   &  1993 Apr 26.92 &  1996.8   &  
 87.75 &   0.360   & --12.42\nl    
N~67   &  00:58:37.3  & --71:35:49  & 
 F501N   &  1993 Jan 09.85  & 1996.9   &    
83.33    &  0.064  & --12.40\nl    
L~343  &   00:58:42.6   &  --72:57:00   & 
 F501N  &   1993 Jul 10.26 & 1995.8   &   
122.98  &  0.126   & --12.22\nl    
L~536   &  01:24:11.8   &  --74:02:34  
&    F501N  &   1993 Jul 06.66  & 995.8   &    
42.82   &   0.130   &  --12.68\nl  
\cutinhead{LMC}
N~97   &  05:04:51.9   &  --68:39:10  
&     F501N   &  1992 Nov 18.12 &  1803.2  &       389.99  &   0.282   & -11.72\nl  
  &   &   &    F501N  &   1994
Feb 02.81   & 1995.9 &  &  & --11.77\nl  
  &   &   &  F501N\tablenotemark{b}  & 1994 Feb 02.84   &   995.9  &    & & --11.73\nl
N~24  &  
05:06:09.3   &  --67:45:29  &    F501N   &  1992 Dec 12.94   &
996.9   &     328.53   &  0.449   & --11.79\nl 
 N~192  &  
05:09:37.3   &  --70:49:09   &  F501N   &  1993 Mar 03.01  &
996.8   &       358.85& 0.145&  --11.74\nl 
  &  &   & 
  F501N   &  1994 Feb 06.81  & 1995.9  &  &  & --11.77\nl 
&  &   &  F501N\tablenotemark{b}   &  1994 Feb 06.84  &  995.9  
&  & & --11.74\nl 
WS~12  &  05:10:50.0   & --65:29:31  &   F501N   &  1993 Apr
28.86  & 1996.8  &  392.38&              0.118  &--11.72\nl
LM~1--27  &   05:19:20.7  &  --66:58:07   &     F501N   &  1993 Jun
20.95 &  1996.8  &     187.76 &   0.049   &   --12.02\nl   
%
%N~201   &  05:24:55.1   &  --71:32:56  &   F486N  &  1991 Jun
%27.87 & 836.4  &  339.30  &              --12.310   &  0.218   &   123.06\nl 
%
%  &  &    &  F501N   &  1991 Jun 27.81   &  734.3  &  339.30   & 
%--11.261  & 0.949  & 988.60\nl   
%
N~52   &   05:28:41.2   &  --67:33:39  & 
 F501N   &  1992 Nov 18.91 & 996.9   &         
 318.94 &   0.220  & --11.79\nl    
%
%N~66  &   05:36:20.8   &  --67:18:08   &
%  F486N  & 1991 Jun 27.00  & 974.3 &           --12.650 
% &  0.043  & 57.58\\   
%
N~66&    05:36:20.8  &    --67:18:08  & 
  F501N   &  1991 Jun 26.94  &
  540.3  & 365.20 & 0.130  & --11.74\nl    
&   &    &    F501N   &  1993 Jul 10.19  &   1995.8  &  325.94 &  0.247   &
--11.81\nl   
&   &     &    F501N   & 1994 Feb 05.30  &   1995.6  &  
 & & --11.80\nl 

   &   &    &   F501N\tablenotemark{b}  &   1994 Feb 06.33  &   995.9   & 
 & & --11.76\nl    
LM~1--61 &   06:10:25.5  &   --67:56:21  &  F501N  &  1992 Nov 18.84 &  1778.3  &    281.64    &   0.100  
&  --11.85\nl

\enddata

\tablenotetext{a} 
{Coordinates: STScI Guide Star Selection System}

\tablenotetext{b}{these filters are really F501N+F1ND}

\end{deluxetable}

\clearpage

\begin{deluxetable}{l r l l c c r}
\scriptsize

\tablecaption{Morphology, Diameters, and Optical Depths}

\tablehead{
\colhead{Name}&   \colhead{SMP}&  \colhead{camera}&  \colhead{Morph.\tablenotemark{a}}&  
\colhead{$\theta$}& \colhead{D} & \colhead{opt. depth}\nl
&&&& [arcsec]& [pc]& \nl
}
\startdata
\cutinhead{SMC}
N~1    &1  &PC1  &R & .241  &     .068      &     thin\nl
N~2\tablenotemark{b}    &2  &FOC  &E(Es)&      .637 &      .180 &          thin\nl
N~4    &3  &FOC,PC1 &B(Bbc) & 2.64&      .747&   thin   \nl
N~5    &5  &FOC  &R & .621  &   .176&           thick \nl
N~6$^b$    &6  &PC1  &R  & .304&       .086 & thin   \nl
N~18   &10 &FOC  &B(Bbc) & 2.64  &  .747 &   \nodata \nl
L~305\tablenotemark{b}  &21 &FOC  &E(Es) & 3.00 &     .846&     thin\nl
N~67\tablenotemark{b}   &22 &FOC  &B(Bbc) &2.74   &   .774&      thick\nl
L~343  &23 &FOC  &B(Bbc)  & 2.61 &     .738&     \nodata\nl
L~536\tablenotemark{b}   &28 &FOC,PC1 &E&  3.50  &    .990&      thin \nl
\cutinhead{LMC}
\nodata& 2& PC1& R& .408&         .100 &     thick\nl
N~78  & 8  &PC1  &R & \nodata& \nodata   &  thin\nl
LM~2--5\tablenotemark{b} & 20 &PC1  &B(Bbc)& .823  &     .202 & thick \nl
N~97\tablenotemark{b}\tablenotemark{c}   
& 21 &FOC  &Q   &1.15    &  .281& thick \nl
N~24  & 23 &FOC  &R    &\tablenotemark{d}  &   \tablenotemark{d}& thin    \nl
N~192\tablenotemark{c} & 32 &FOC  &R & 1.083  &    .266& thick \nl
WS~12 & 35 &FOC,PC1 &E(inc)&   1.59  &    .391   & thick\nl
WS~16 & 40 &PC1  & E(Es)& .783&      .192 &          thick\nl
LM~1--27& 45 &FOC & E(inc)&  2.36 &     .578&     thick\nl
N~122\tablenotemark{b} & 47 &PC1  & E? &  .412&       .101 &  thin\nl
%N~201\tablenotemark{b} & 62 &1  & E& \nodata& \nodata& \nodata   \nl
N~52   &66 &FOC  & E(Es)& 1.08   &   .266 &     \nodata\nl
\nodata &72 &PC1 & B & \nodata& & thin \nl
N~60   &76 &PC1  & R   & \nodata& \nodata& thin \nl
N~69   &85 &PC1  & R    & \nodata& \nodata& thin  \nl
N~215\tablenotemark{b}  &87 &PC1  & B(Bbc)  & 1.01 &      .248    &  thick \nl
\nodata\tablenotemark{b}&96 &PC1 & B(Bbc) & .905 &      .222    &  thick\nl
LM~1--61 &97 &FOC  & R(Rs) &1.18  &    .289    &      thin\nl
\enddata

\tablenotetext{a}{R=Round, E=Elliptical, B=Bipolar, Q=Quadrupolar, Rs=Round with (inner) structures, Es=Elliptical with (inner) structures, 
Einc=Elliptical incomplete, Bbc=Bipolar core} 
\tablenotetext{b}{Type I PN}
\tablenotetext{c}{pre-- and post--COSTAR FOC images}
\tablenotetext{d}{see text, $\S$ 3.2.}

\end{deluxetable}

\clearpage

\begin{deluxetable}{l    l    l   r  }
\scriptsize

\tablewidth{10truecm}
\tablecaption{Ellipticities, Expansion Velocities, and Dynamical Times}

\tablehead{
\colhead{Name}&   \colhead{q} & \colhead{V$_{\rm exp}$}&  
 \colhead{$\tau_{\rm dyn,yr}$} \nl
& & [km s$^{-1}$]& [10$^3$ yr]\nl}  

\startdata

\cutinhead{SMC}
N~1 &  1.0   &8.46 & 3.94      \nl
N~2  & .92   &17.7 & 4.98  \nl
N~4  & .69   &18.1 &  20.2      \nl
N~5  & 1.0   &16.0 &5.38   \nl
N~6  & 1.0   &19.3\tablenotemark{a} &2.18        \nl
N~18 & .80   &13.2 & 27.7    \nl
L~305 & .90  &19.3 & 21.4     \nl
N~67  & .71  &27.9 & 13.6      \nl
L~343  &.94   &17.4 & 20.8      \nl
L~536  &.80  &29.3 & 16.5   \nl
\cutinhead{LMC}
SMP~2 &  1.0&5.44 & 9.01  \nl
N~78 &1.0   &13.9 & \nodata  \nl
LM~2--5 &1.0 &14.2& 6.98  \nl
N~97  &1.0  &27.0 &5.11 \nl
N~24  &1.0  &11.5 &\nodata   \nl
N~192 &1.0  &23.2 &5.60  \nl
WS~12 &.82  &22.7 &8.44   \nl
WS~16 &.64  &30.0 &3.14   \nl
LM~1--27&.75 &20.2 &14.0   \nl
N~122 &.50  &43.2 &1.14  \nl
N~52  &.85  &12.7 &10.3   \nl
SMP~72 & .64&\nodata  &   \nodata      \nl
N~60 &1.0   &15.9 & \nodata   \nl
N~69 &1.0   &6.21 & \nodata  \nl
N~215&.57   &20.6 &5.91\nl
SMP~96 &.34 &33.5 &3.25   \nl
LM~1--61&1.0 &25.3 &5.60  \nl
\enddata
\tablenotetext{a}{two components: 12.1 and 26.74 km s$^{-1}$}

\end{deluxetable}

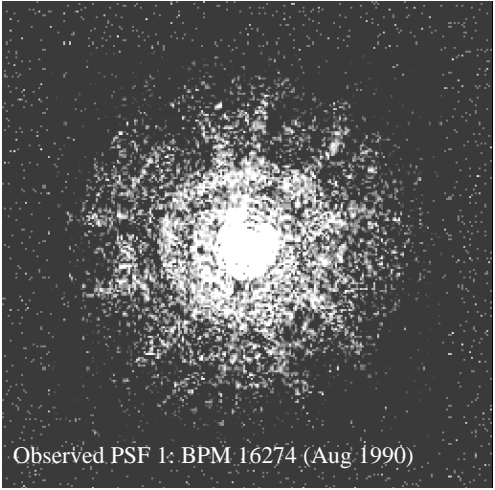
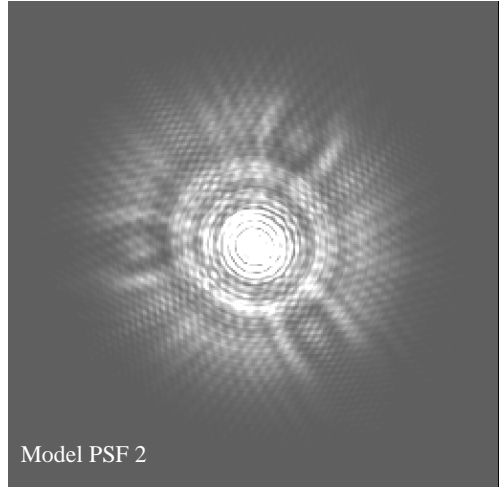
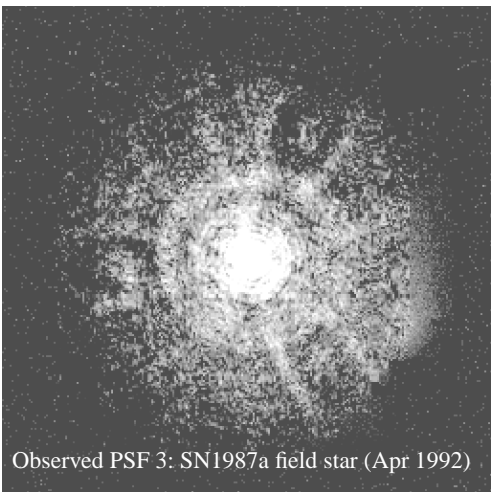
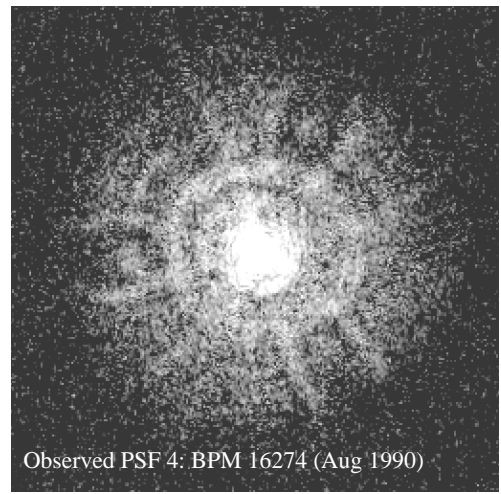
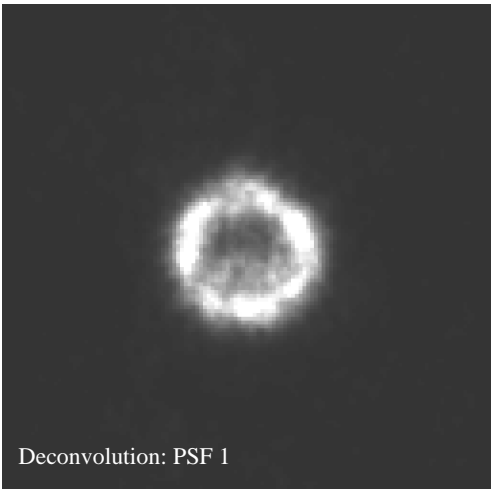
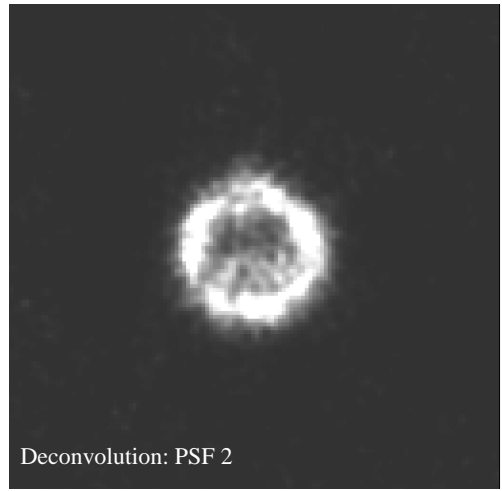
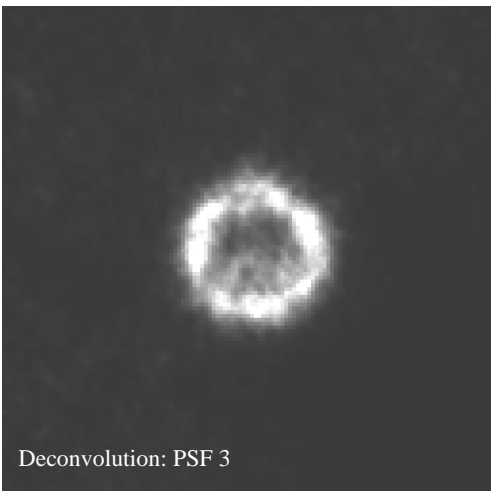
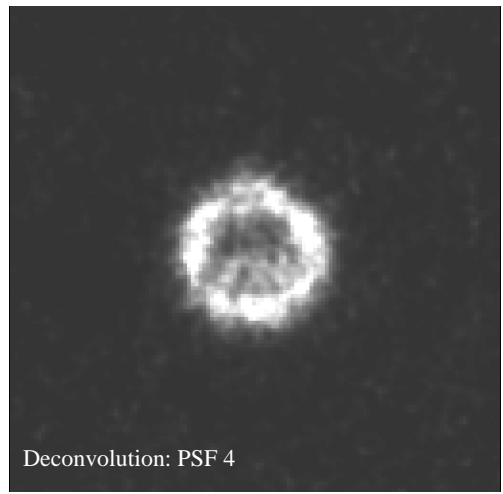
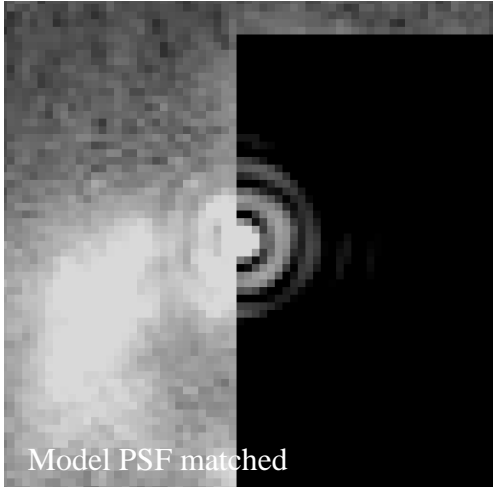
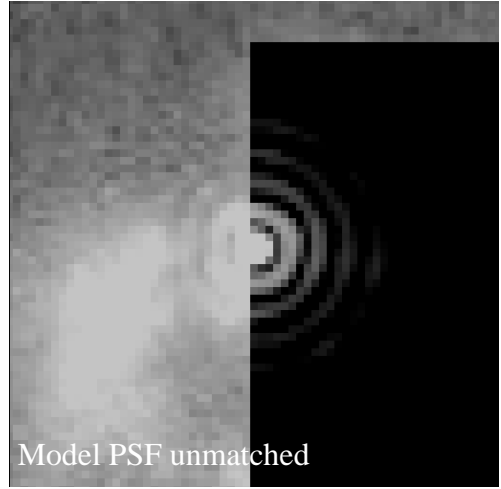
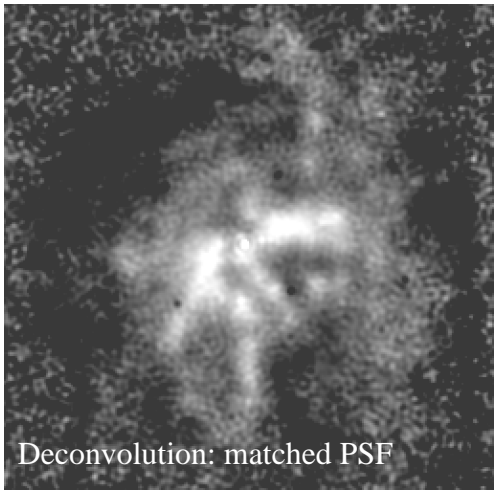
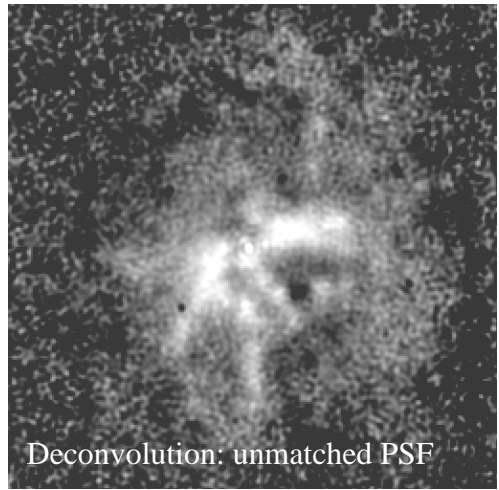
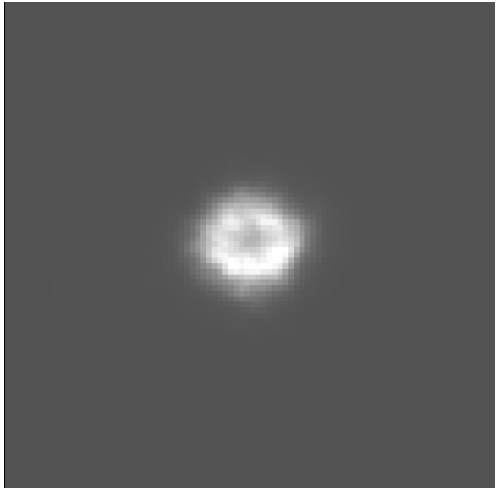
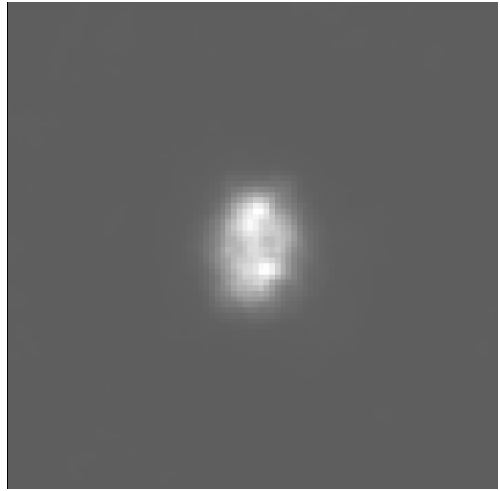
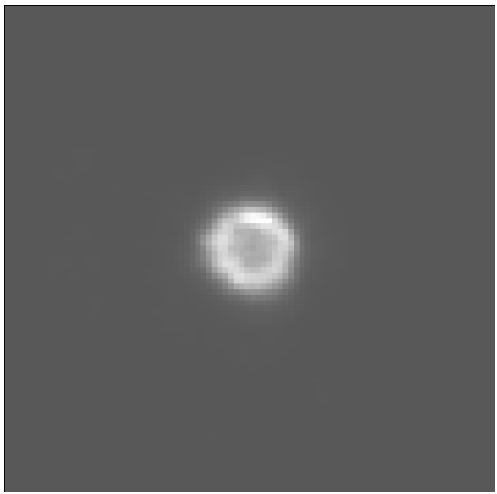
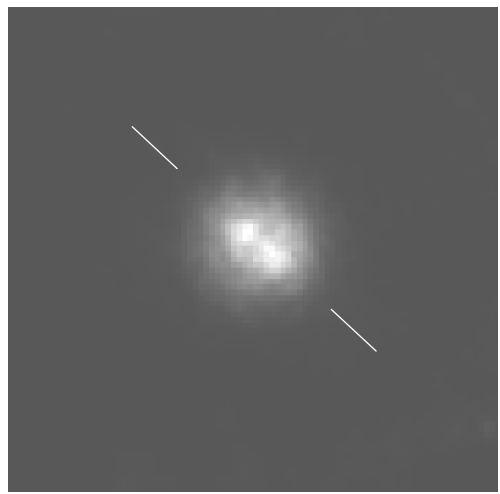
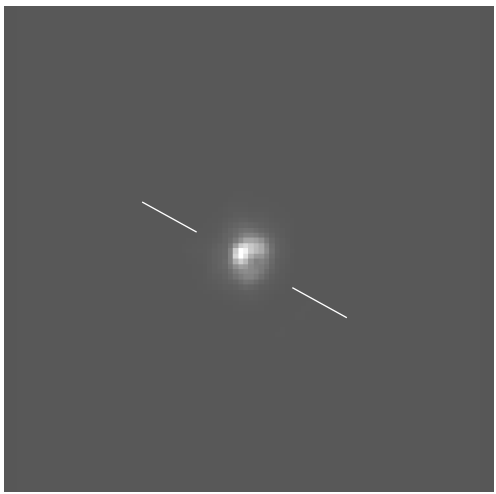
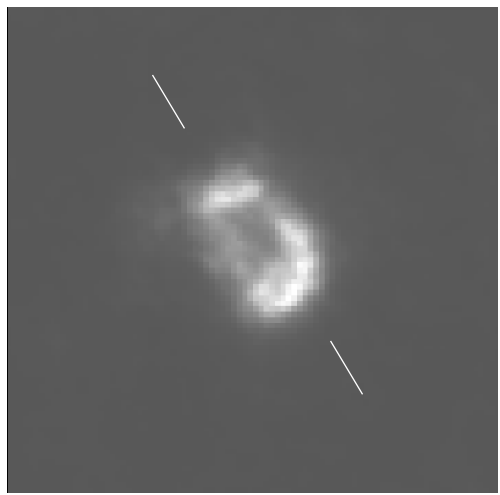
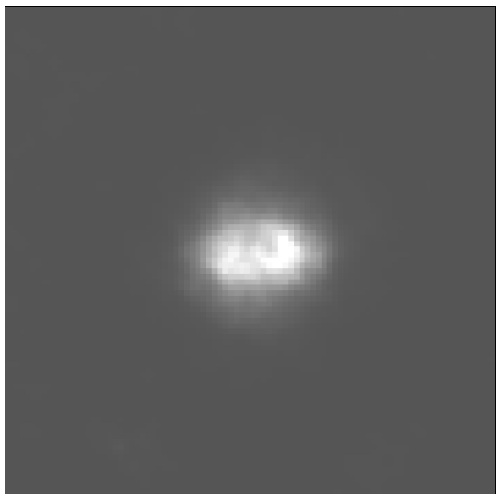
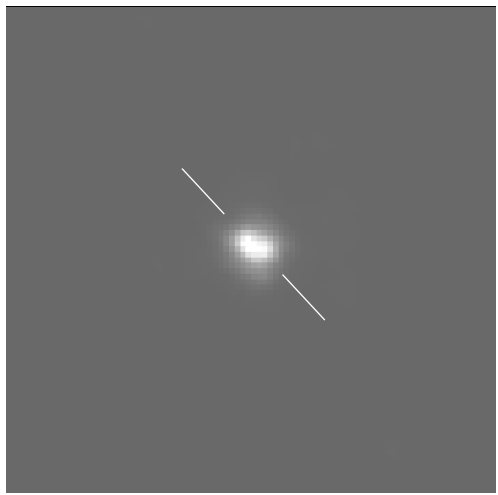
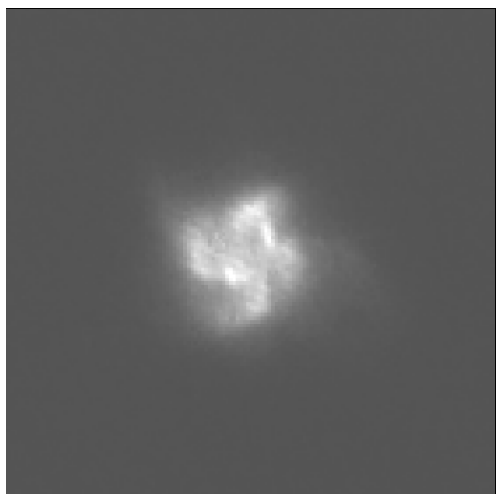
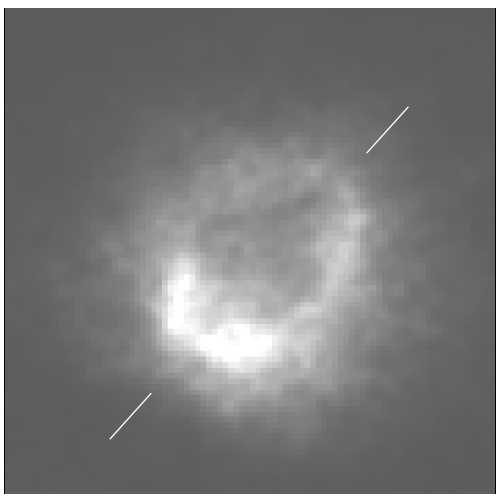
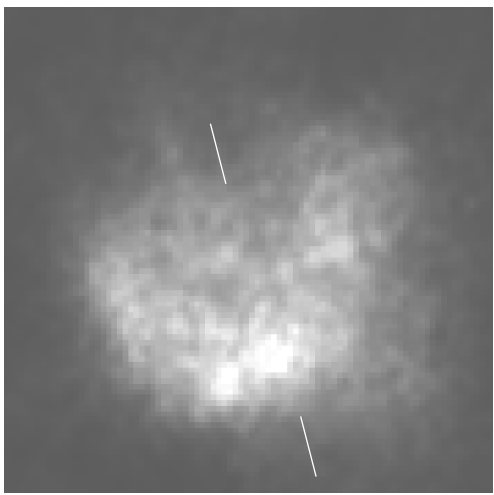
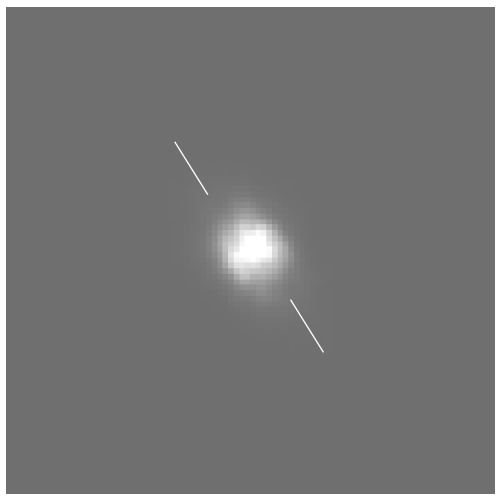
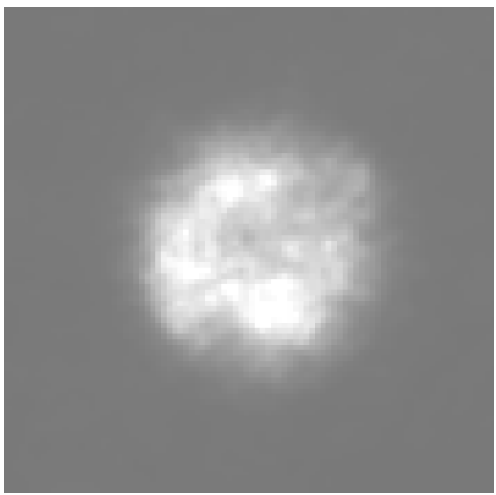
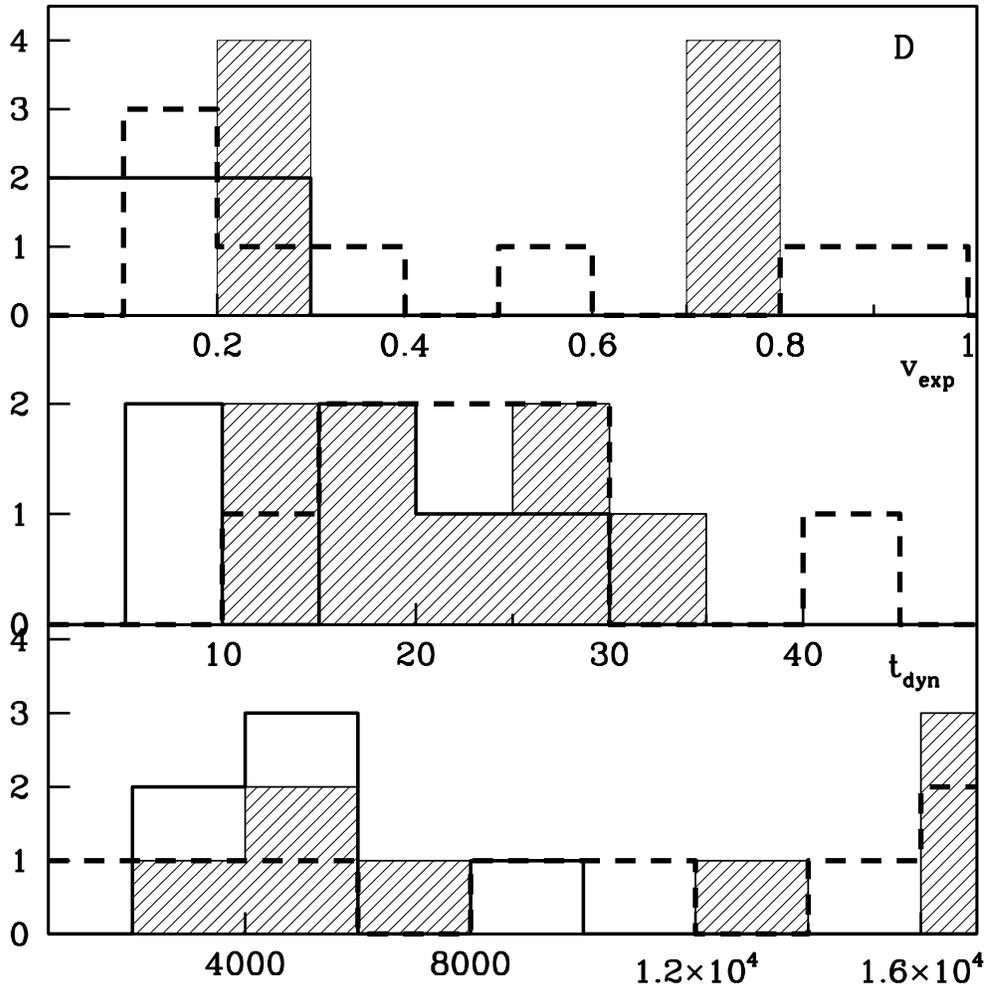
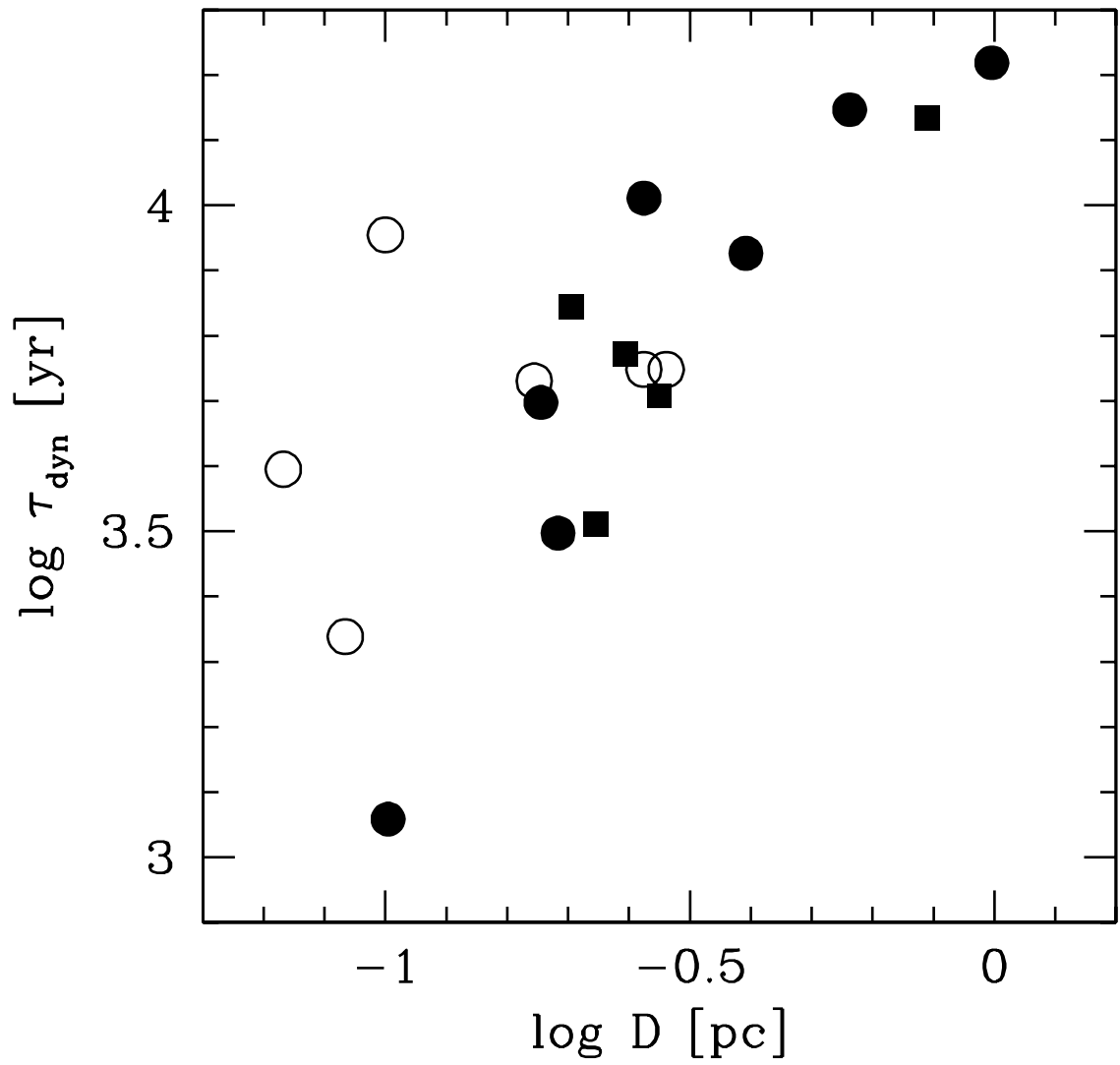
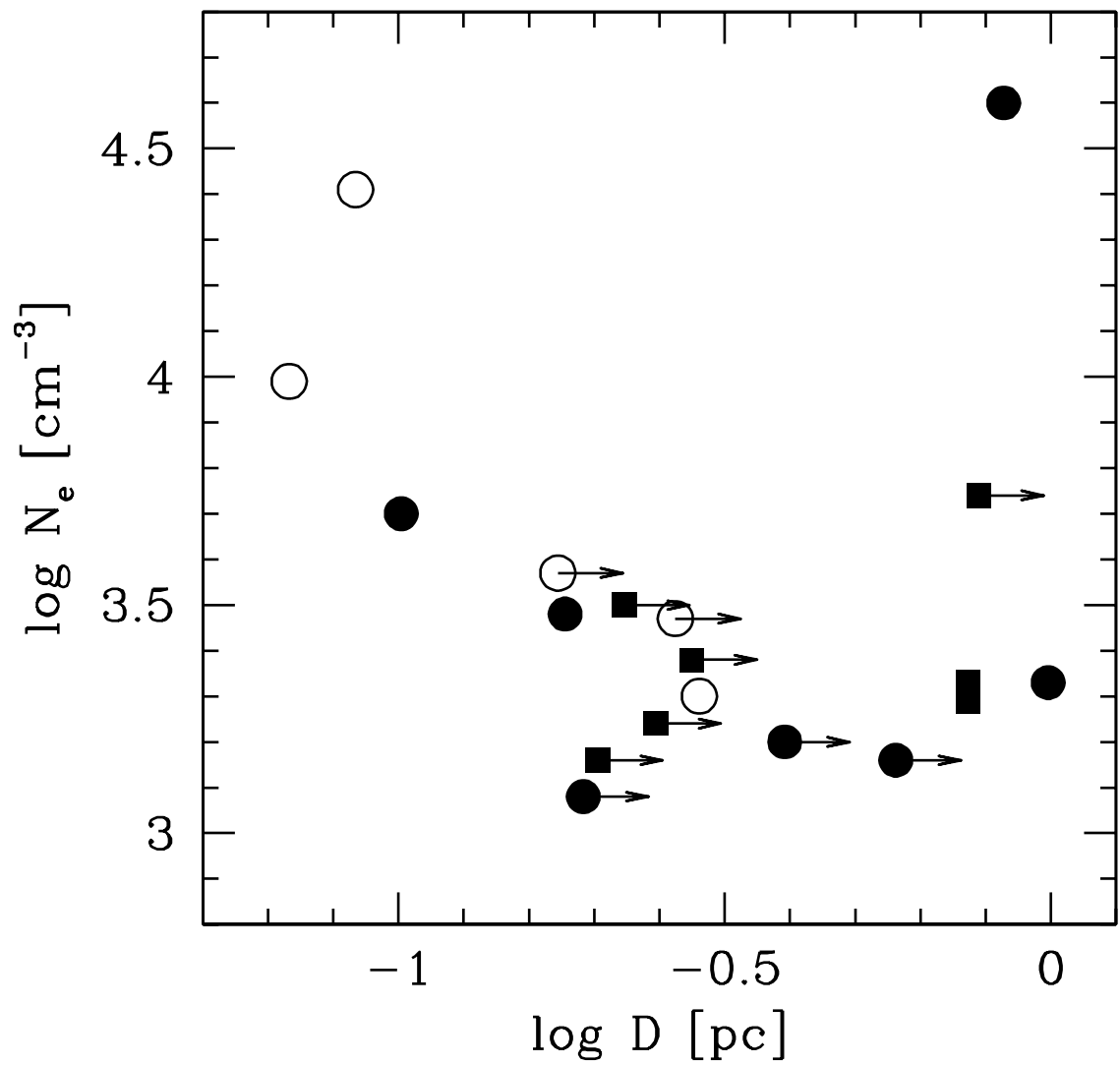
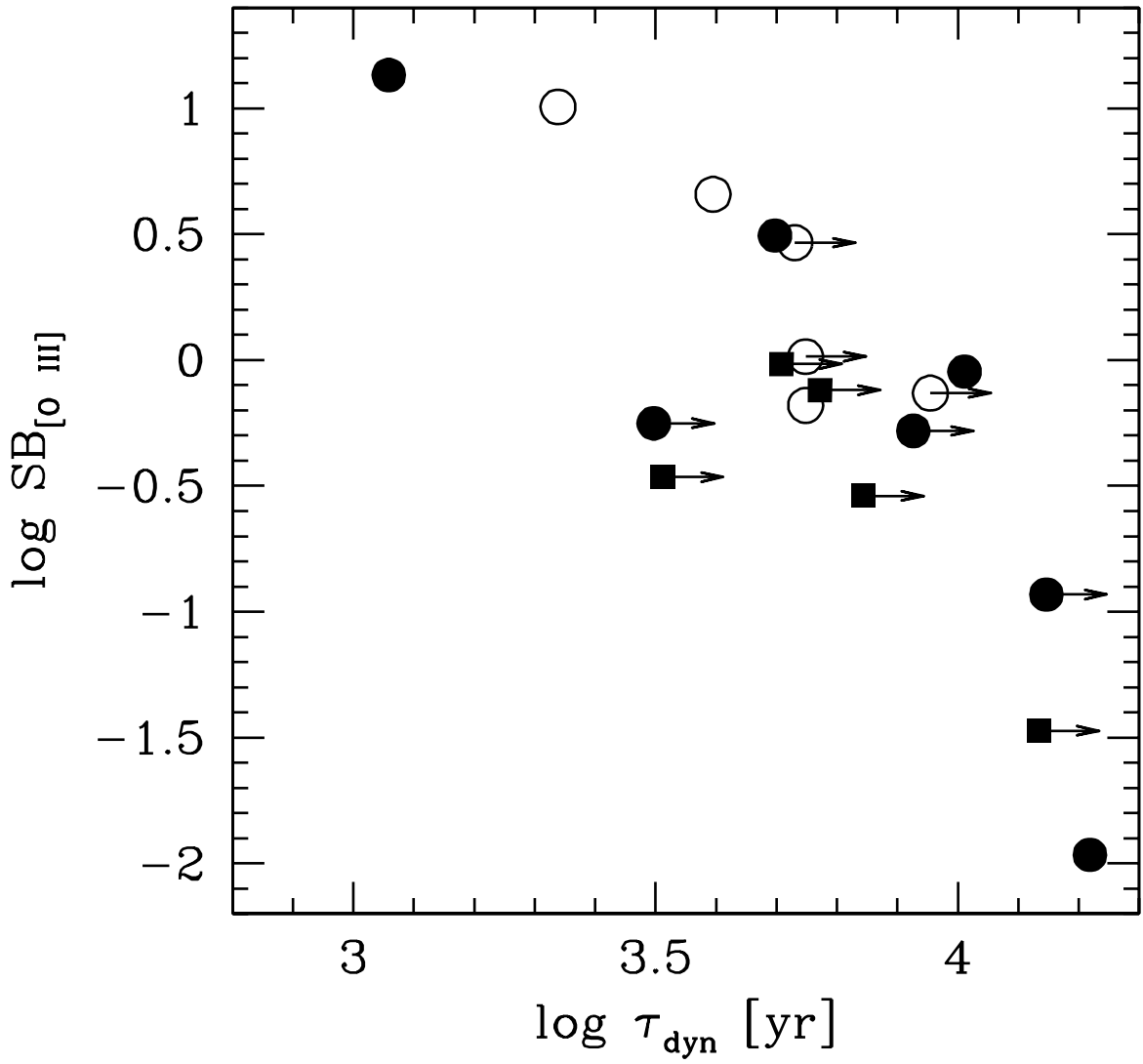
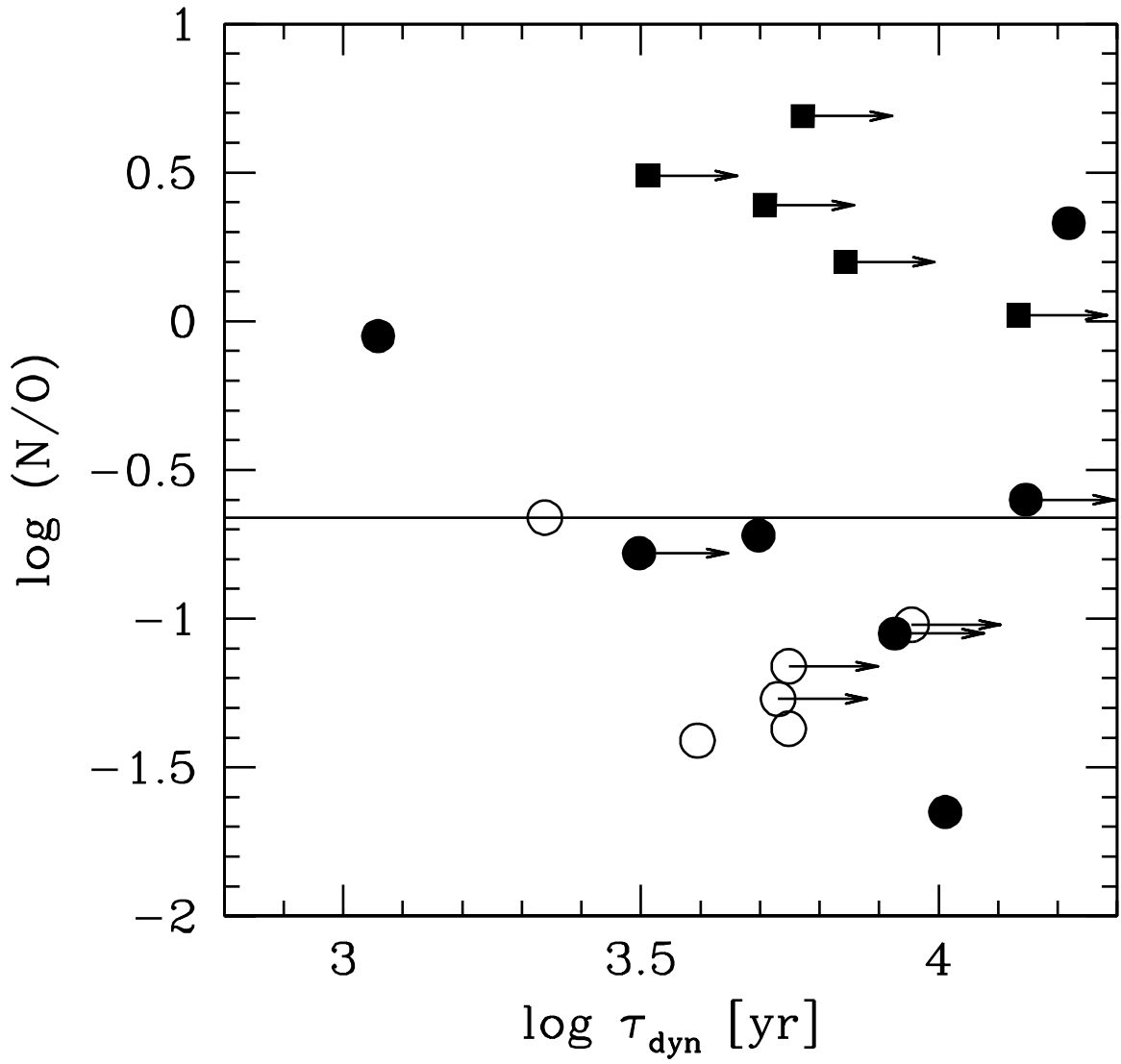
\end{document}